\documentclass[%
superscriptaddress,
frontmatterverbose, 
showpacs,preprintnumbers,
showkeys,
twocolumn,
 amsmath,amssymb,
 aps,
prl,
]{revtex4-1}

\usepackage[pdftex]{graphicx}
\usepackage{amssymb,amsfonts,amsmath}
\usepackage{color}
\usepackage[usenames,dvipsnames]{xcolor}
\usepackage{sidecap}

\definecolor{darkchocolate}{rgb}{0.55, 0.27, 0.07}

\begin{document}

\title{Translationally invariant colloidal crystal templates}

\author{Pankaj Popli}
\affiliation{TIFR Centre for Interdisciplinary Sciences, 36/P Gopanpally, Hyderabad 500107, India}
\author{Saswati Ganguly}
\affiliation{Institut f\"ur Theoretische Physik II: Weiche Materie, Heinrich Heine-Universit\"at D\"usseldorf, Universit\"atsstra{\ss}e 1, 40225 D\"usseldorf, Germany}
\author{Surajit Sengupta}
\affiliation{TIFR Centre for Interdisciplinary Sciences, 36/P Gopanpally, Hyderabad 500107, India}

\date{\today}


\begin{abstract}
We show that dynamic, feed-back controlled optical traps, whose positions depend on the instantaneous local configuration of particles in a pre-determined way, can stabilise colloidal particles in finite lattices of {\em any} given symmetry. Unlike in a static template, the crystal so formed is translationally invariant and retains all possible zero energy modes. We demonstrate this {\it in-silico} by stabilising the unstable two-dimensional {\em square} lattice in a model soft solid with isotropic interactions.  
\end{abstract}

%
%
\maketitle

Discovery and development of techniques to form stable, complex assemblies of colloidal particles has emerged as a rather vibrant sub-field of soft condensed matter physics and materials science~\cite{colbks1,colbks2,colbks3,colbks4}. Colloidal particles, synthesized using a variety of routes into an array of shapes and sizes from nanometers to microns are now, readily available for use~\cite{colbks4}. Apart from many technological applications, assemblies of such colloids into ordered crystals offer us unique insights into properties of ordinary solids and their behaviour. This is facilitated by the relatively large size and consequent slow timescales ~\cite{phonons1,phonons2,phonons3} of a colloidal particle enabling one to use simple optical means to observe and manipulate them~\cite{bowick,video,defects}. 

One encounters, mainly, two paradigms related to the assembly of colloids into complex structures. In the first case, the interactions between colloidal particles are tuned, either by controlling the shape or by adding specially reactive patches or tethers~\cite{patchy1,patchy2}. While one obtains a lot of control over the symmetry and properties of the colloidal crystals so produced, the lattice is fixed once the interactions are. On the other hand, one may also produce ordered colloidal crystals using templates~\cite{anton}. These templates may be either permanent, such as etched onto a surface, or reconfigurable, if produced by optical means~\cite{laser1,laser2,laser3}. The latter technique has the advantage that a large variety of crystal~\cite{laser1}, quasi-crystal~\cite{laser2} and even random structures~\cite{laser3} may be produced. Nevertheless, periodic crystals induced by static templates necessarily suffer from a fundamental flaw - templates break translational invariance~\cite{laser4,CL}. Uniform translations now cost energy and one or more zero modes of the crystal become massive making the spectrum of vibrational frequencies resemble that of a trivial Einstein crystal~\cite{ashcroft}. While for many applications, such as colloidal epitaxy~\cite{anton, colbks4}, this may not present a problem, for many others it is an issue to be addressed. For example, constructing colloidal models of solid-solid transformations~\cite{Yodh} and interfaces~\cite{Meyers}, mechanical behaviour of crystals~\cite{rob}, crystal - glass transition~\cite{devit} etc. require that spurious effects from such explicit spatial symmetry breaking be avoided.    
\begin{figure}[h!]
\begin{center}
\includegraphics[width=0.4\textwidth]{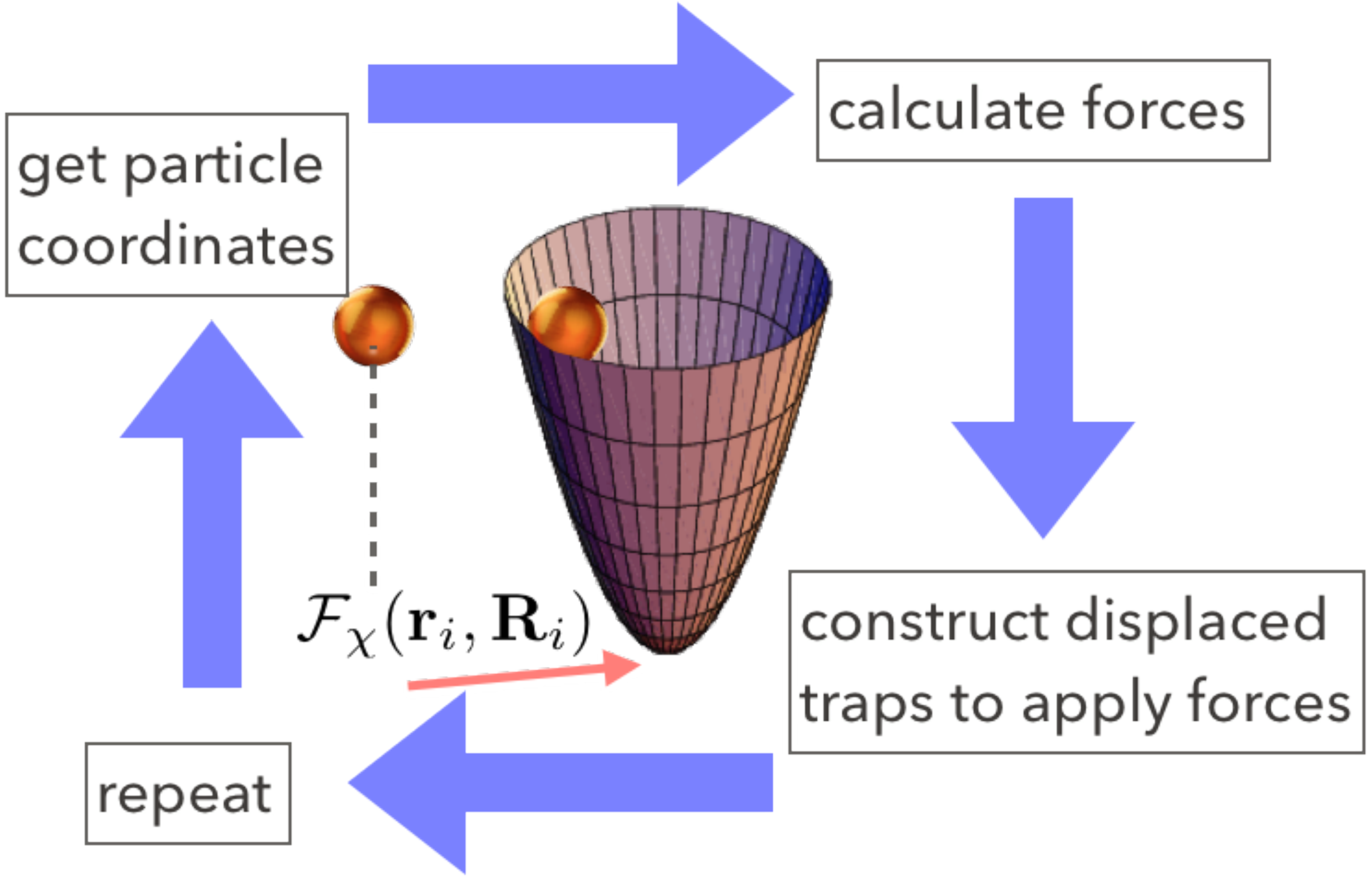}
\caption{\label{schematic}Schematic diagram showing the steps involved in generating feed-back controlled optical traps. The particle coordinates are obtained from video microscopy. The neighborhood of each particle is recorded and the force ${\cal F}_\chi({\bf r}_i,{\bf R}_i)$ (see text) calculated from the Hamiltonian~\ref{hamil}. Next, the positions of optical traps needed for applying these forces are calculated and the traps deployed. If this procedure is continuously repeated for all particles at a rate comparable to typical vibrational frequencies of the colloidal particles, a uniform stabilising field is obtained.}
\end{center}
\vskip -.3cm
\end{figure}

In this Letter, we propose a means to stabilize translationally invariant colloidal crystals in any desired lattice symmetry. The particles may, in principle, have almost any kind of pairwise or many-body interactions. For example, open lattices such as the square or honeycomb in two dimensions, which are unstable for most colloids interacting with central, pair potentials may now be stabilised. Our proposed method is described below.

First, the target lattice $S_T$, is ``read in'' as a set of reference coordinates $\{{\bf R}_i\}$ (see Fig.~\ref{schematic}). Note that there is no restriction on the choice of $\{{\bf R}_i\}$. Next, around each particle $i$, we fix a neighborhood $\Omega$  containing a fixed set of tagged particles whose instantaneous coordinates ${\bf r}_i$ are recorded. Particle positions undergo thermal fluctuations due to the presence of the solvent. Using the well defined projection formalism, which we have established in previous publications~\cite{sas1,sas2}, we separate these thermal particle displacements into affine and non-affine subspaces. A laser tweezer is used to exert additional forces, ${\cal F}_\chi({\bf r}_i,{\bf R}_i)$ to particle $i$ which bias displacement fluctuations so that the non-affine component of the displacements is suppressed. Since  ${\cal F}_\chi$ depends on {\em instantaneous} particle positions, they need to be  be continuously updated by tracking particle trajectories in real time. Since intrinsic timescales of colloids are large, this is achievable using current video microscopic and spatial light modulation technology~\cite{HOT}. We derive below ${\cal F}_\chi$ for any given $S_T$. 
 
The additional forces are computed from the following extended Hamiltonian~\cite{sas2,sas4}, ${\cal H} = {\cal H}_0 + {\cal H}_X$. Here ${\cal H}_0$ represents {\em any} Hamiltonian for interacting particles and,   
 \begin{equation} 
{\cal H}_X = -  h_X \sum_i^N \sum_{jk\in \Omega} ({\bf u}_j-{\bf u}_i)^{\rm T}{\bf P}_{j-i,k-i}(\{{\bf R}_i\})({\bf u}_k-{\bf u}_i).
\label{hamil}
\end{equation}

${\cal H}_X$ involves the ``projection operator" ${\mathsf P}$ and the particle displacements ${\bf u}_i = {\bf r}_i - {\bf R}_i$. The projection operator is a function only of the reference lattice and is given by ${\mathsf P}^2 = {\mathsf P} = {\mathsf I}-{\mathsf R}({\mathsf R}^{\rm T}{\mathsf R})^{-1}{\mathsf R}^{\rm T}$ and ${\mathsf R}_{j}^{\alpha,\gamma\gamma^{\prime}} = \delta_{\alpha\gamma}R_{j}^{\gamma^{\prime}}$, where the greek indices $\alpha = 1,2$ go over spatial components. 

Note that (\ref{hamil}) preserves translational invariance viz., ${\bf u}_i \to {\bf u}_i +{\rm constant}$. 
One can also show~\cite{sas2}, that ${\cal H}_X = -h_X \sum_i \chi_i$ where $\chi_i$ is the least square error made replacing particle displacements in $\Omega$ by the ``best fit'' {\em affine} strains~\cite{falk}. The quantity $\chi_i = \Delta^{\rm T} {\mathsf P} \Delta$ where $\Delta$ is the column vector of displacement differences~\cite{sas1} with components $\Delta_j = {\bf u}_j - {\bf u}_i$ between particles $i$ and all its neighbours $j$ within $\Omega$. The projection operator therefore projects out the {\em non}-affine part of $\Delta$ and $\chi_i$ is the local non-affine parameter. Finally, the forces ${\cal F}^i_\chi = -\partial {\cal H}_X / \partial {\bf u}_i$. We show below that {\em suppressing non-affine fluctuations using negative values of $h_X$ stabilises $S_T$}.    

\begin{figure}[ht!]
\begin{center}
\includegraphics[width=0.5\textwidth]{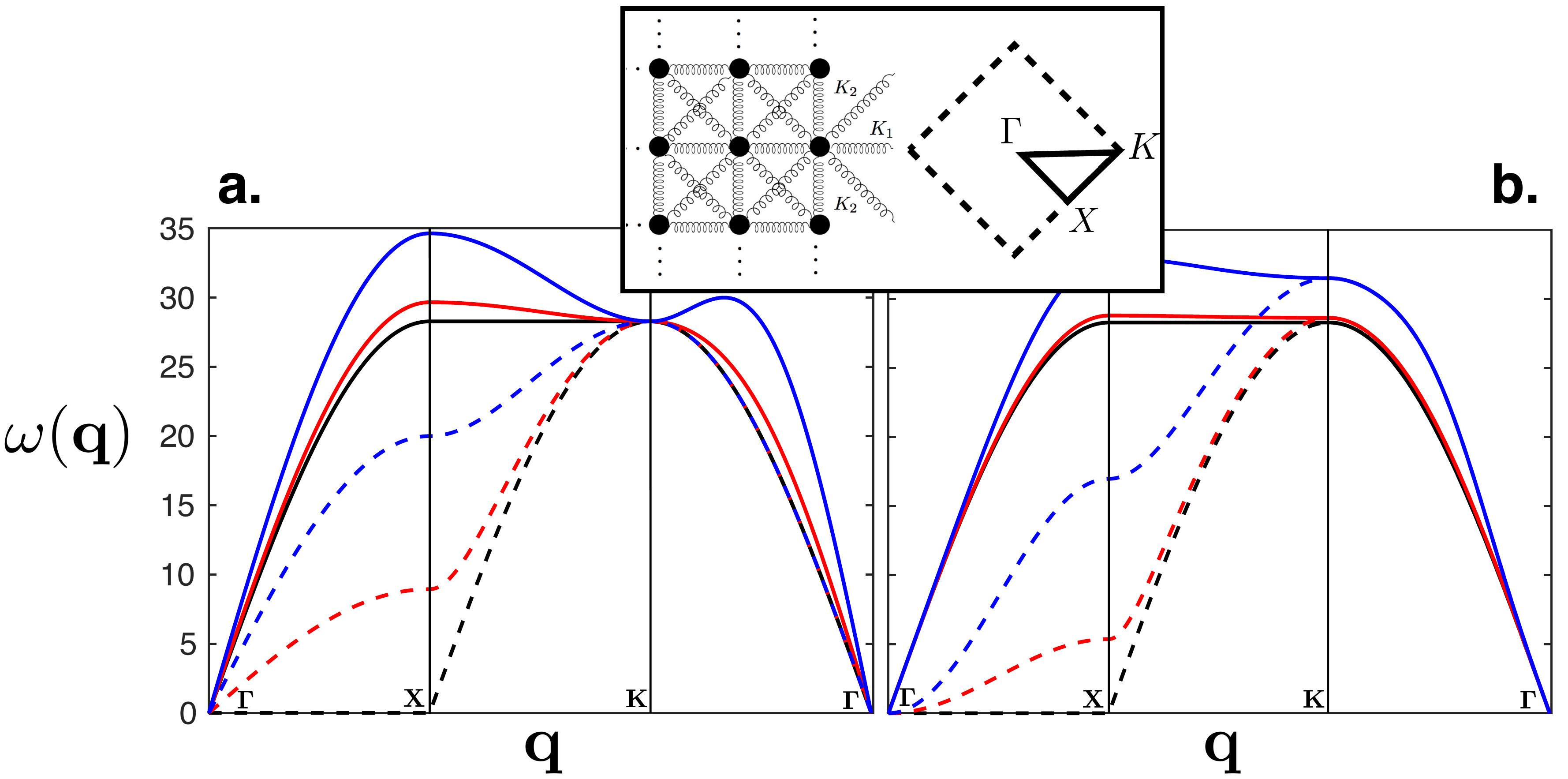}
\caption{\label{dqhar}{\bf a.} Phonon dispersion curve $\omega(q)$ plotted along high symmetry directions for a square lattice network of vertices connected by springs. Both nearest neighbour and next nearest neighbour springs with spring constants $K_1, K_2 > 0$  are needed for stability. The longitudinal (transverse) modes are shown with solid (dashed) lines. The colours denote $K_2 = 0$ (black), $0.1$ (red) and $0.5$ (blue). {\bf b.} Phonon dispersion for  $K_2 = 0$ but now for various values of $h_X = 0$ (black), $0.003$ (red), $0.03$ (blue).  Insets show the interaction volume $\Omega$ in the square lattice with bonds (left) and the high symmetry points of the corresponding Brillouin zone (right)}
\end{center}
\vskip -.2cm
\end{figure}

It is instructive  to demonstrate this first for the case of a two dimensional (2d) square lattice of vertices (see Fig.~\ref{dqhar}). The nearest and next nearest neighbour vertices are connected by harmonic springs of strengths $K_1 = 1$ and $K_2$ respectively. The square lattice is mechanically unstable in the limit $K_2 \to 0$ due to softening of transverse phonon modes~\cite{spring}. This is illustrated in Fig~\ref{dqhar} {\bf a} where we have plotted the phonon dispersion $\omega({\bf q})$ for this harmonic square network for three different values of $K_2$. For $K_2 = 0$ the transverse mode disappears. We now add the term proportional to $h_X$ in ~(\ref{hamil}) to the harmonic hamiltonian ${\mathcal H}_0$ using the interaction volume, namely, the first and second neighbour shells of the square lattice as our choice of $\Omega$ (see Fig.~\ref{dqhar} inset). The dynamical matrix~\cite{CL,ashcroft}, corresponding to the full Hamiltonian ${\mathcal H}$ can be written as ${\mathcal D}^{\mu \nu} = {\mathcal D}^{\mu \nu}_0 + {\mathcal D}^{\mu \nu}_X$, where ${\mathcal D}^{\mu \nu}_0$ is the dynamical matrix from ${\cal H}_0$ and 
\begin{equation}
\mathcal{D}_{X}^{\mu\nu} = -\delta^{\mu\nu}h_X\sum_{\bf R}\frac{\partial^{2}\chi}{\partial u({\bf R})^{\mu}\partial u(0)^{\nu}}e^{-i {\bf q}\cdot{\bf R}}
\end{equation}
with the lattice sum over the reference set $\{{\bf R}_i\}$. After completing the sum over the square lattice, we obtain $\mathcal{D}_{X}^{\mu\nu}  = 2h_X\mathcal{A}_{X} \delta^{\mu\nu}$. 
The function ${\mathcal A}_X (\bf q) \sim q^4$ for small wave-numbers, so that ${\mathcal D}_X$ does not affect the elastic properties, e.g. the speed of sound, of a mechanically stable lattice. Further, these results do {\em not} depend on the nature of  ${\cal H}_0$.

In Fig.~\ref{dqhar}{\bf b} we plot the resulting phonon dispersion curves all at $K_2 = 0$ but for three different values of $h_X$. As $h_X$ is made more negative, the transverse phonon mode appears to revive. One must note however that as $q \to 0$ the nature of $A_X({\bf q})$ dictates that the speed of transverse sound vanishes in the hydrodynamic limit. Nevertheless, for finite lattices, this limit is never reached since wave-numbers are cut off at $q = 2 \pi /L$ where $L \propto \sqrt{N}$ is the linear system size. 

Consider, next, a system of particles interacting by the soft, purely repulsive, ``Gaussian core" model potential (GCM)~\cite{gaucor1,gaucor2} viz. $\phi_{ij} = \epsilon \exp(-r_{ij}^2/\sigma^2)$. The parameters $\epsilon$ and $\sigma$ set the energy and length scales respectively and may be taken as unity. The GCM has been used to describe interacting star polymers which form a number of interesting solid phases in three dimensions. In 2d, this system freezes into a triangular lattice with a possible intervening hexatic phase~\cite{gaucor3, zu}. GCM is also useful because the simple form of $\phi_{ij}$ makes many analytic calculations possible.  We study the model at reduced density $\rho = 0.5$ and temperature $ T =  1\times10^{-3}$ where a triangular solid is stable~\cite{zu}. In general, we found that temperature plays a relatively minor role as long as anharmonic effects do not dominate. 
\begin{figure}[h!]
\begin{center}
\includegraphics[width=0.48\textwidth]{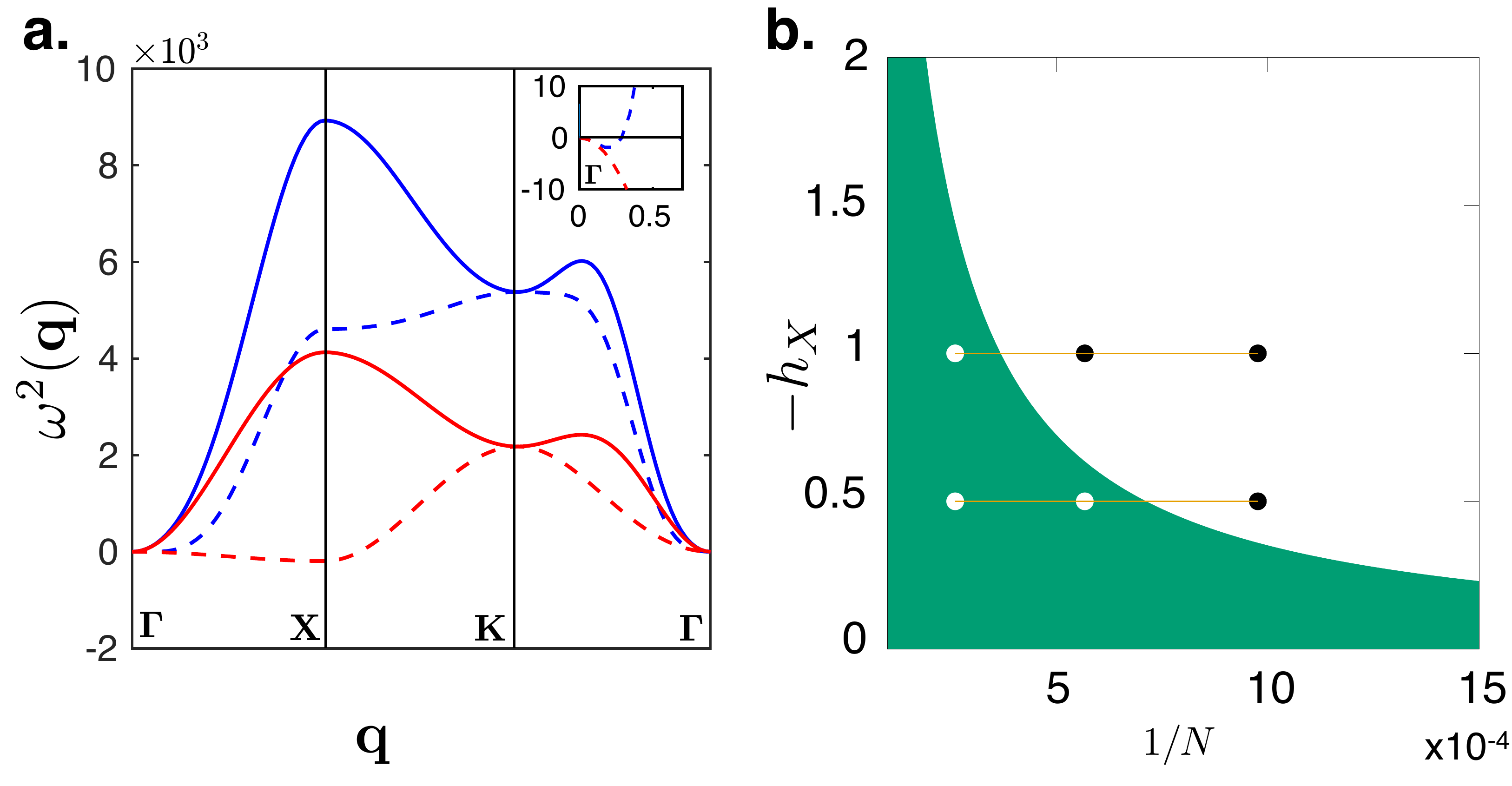}
\caption{\label{GCM}{\bf a.} Phonon dispersion curve $\omega^2(q)$ plotted along high symmetry directions for a square lattice of particles interacting with the GCM potential at $\rho = 0.5$ and $T = 1\times10^{-3}$. The small $q$ region where $\omega^2 < 0$ is shown in the inset. We follow the same convention as in Fig.~\ref{dqhar}. There are two sets of curves for $h_X = 0$ (red curves) and $h_X = 0.5$ (blue curves). {\bf b.} A stability diagram for finite size square lattices constructed from the dispersion curve in {\bf a.} The green shaded region represents $-h_X$ values below which the square lattice becomes unstable ($\omega^2(q) < 0$) for $q < 2\pi/L$. We have also marked six points on the graph such that black (white) circles denote stable (unstable) square lattices. }
\end{center}
\vskip -.5cm
\end{figure}   

When particles interacting with the GCM potential are arranged in a {\em square} lattice, one obtains a mechanically unstable solid. Small displacement fluctuations from the ideal square lattice positions (where forces still vanish due to symmetry) makes the solid deform into the stable triangular structure via soft transverse modes~\cite{spring}. This is clear from the calculated dispersion curve shown in Fig.~\ref{GCM}{\bf a} for $h_X = 0$.

We now turn on $h_X$ defined exactly as in the case of the network. As $h_X$ is decreased below zero, again, we see a revival of the transverse phonon mode. Unlike the network however, for all $h_X$, the transverse mode now has $\omega^2 < 0$ in a small region of $q \to 0$. Making $h_X$ more negative can, nevertheless, restrict this region to extremely small $q$ values which are not accessed by a solid of finite size due to the infra red cut-off discussed earlier. Since within a harmonic approximation, temperature enters only as a pre-factor, this leads to a $T$ independent ``stability diagram'' as shown in Fig~\ref{GCM}{\bf b.}  
\begin{figure*}[t!]
\begin{center}
\includegraphics[width=0.8\textwidth]{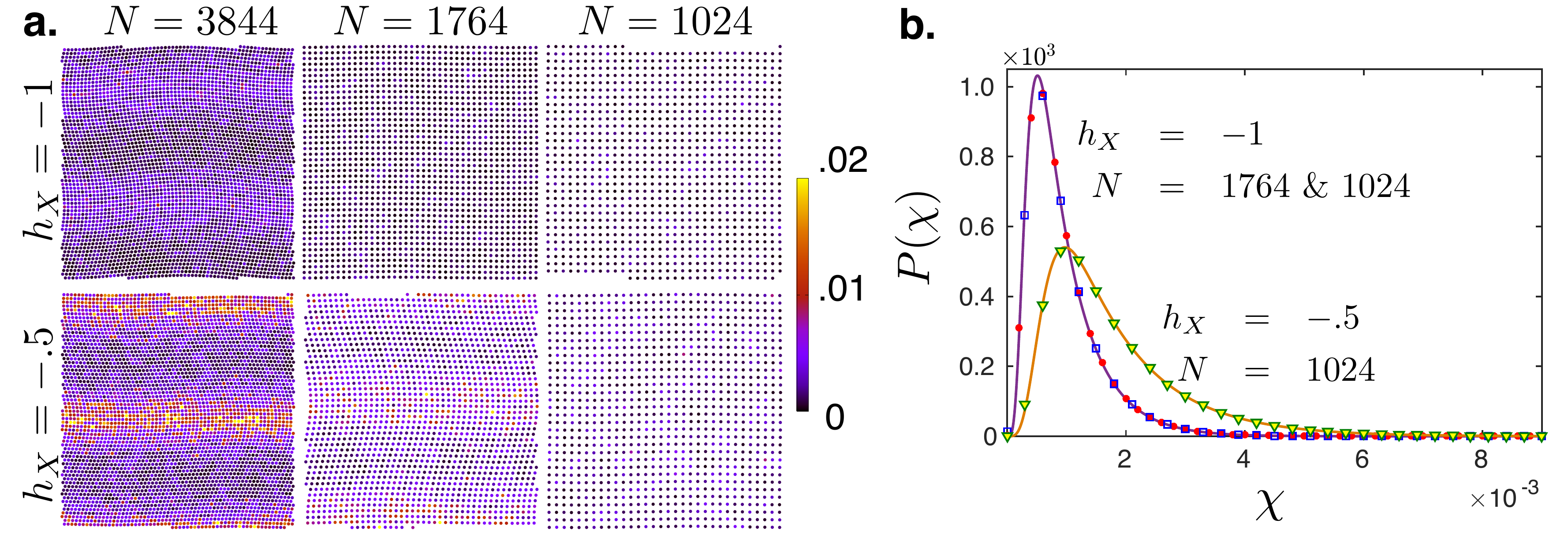}
\caption{\label{config} {\bf a.} Configurations obtained from Monte Carlo simulations after $6\times10^6$ MCS arranged according to the corresponding points marked off in Fig~\ref{GCM}{\bf b} verifying the stability condition. Colours correspond to the local value of $\chi$. {\bf b.} $P(\chi)$ obtained from the equilibrated configurations where the square lattice is stable (symbols). The lines are predictions of the harmonic theory~\cite{sas3}}
\end{center}
\vskip -.5cm
\end{figure*}   

To verify the stability diagram presented in Fig.~\ref{GCM}{\bf b} we perform Monte Carlo simulations with standard Metropolis updates~\cite{frenkel} of a GCM solid with various $N$ and $h_X$ values keeping $\rho$ and $T$ same as before, equilibrating the system for a minimum of $10^{6}$ Monte Carlo steps (MCS) starting from an initial square lattice. To check our results we obtained the probability distribution of the non-affine parameter $P(\chi)$ and compared it with the predictions from harmonic theory using the calculated dynamical matrix ${\cal D}^{\mu \nu}$ as input~\cite{sas1,sas3}.  

Equilibrated configurations from our simulations are presented in Fig.~\ref{config}{\bf a}. The $h_X$ and $N$ values for these configurations are marked in Fig.~\ref{GCM}{\bf b.} It is clear that these results follow the expectations from our stability criterion. Once the stability threshold is breached, the square lattices destabilise due to ${\bf q} \to 0$ modulations. The local non-affine parameter $\chi$ rapidly rises as the crystal becomes unstable and is shown as a colour map. For all the stable square solids studied, our results were indistinguishable from the theoretical prediction (Fig.~\ref{config}{\bf b}). If $\rho$ is increased, the harmonic approximation becomes more accurate and the magnitude of $h_X$ needed to stabilise square lattices becomes even smaller. 
\begin{figure*}[t!]
\begin{center}
\includegraphics[width=.8\textwidth]{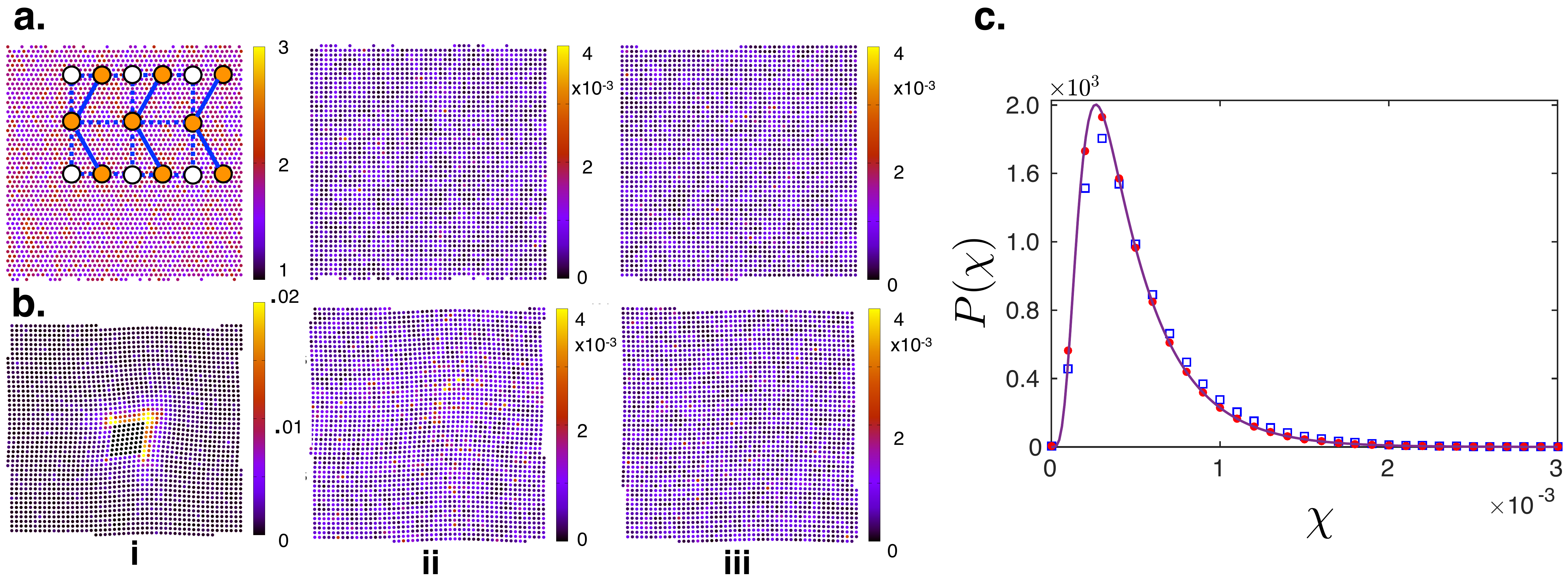}
\caption{\label{patch} Configurations showing the recovery of the square lattice after an initial distortion: {\bf a.} The square lattice of $2500$ GCM particles at $\rho = 0.5$ and $T = 1\times10^{-3}$ was distorted by displacing alternate rows of particles by half a lattice spacing (see schematic inset) producing a triangular lattice close to the $S_E$ structure ({\bf i}). Under a stabilising field of $h_X =-2.0$ the target lattice quickly recovers as seen in {\bf ii} ($4.8\times10^{4}$ MCS) and {\bf iii} ($24\times10^4$ MCS). {\bf b} Here we produce a patch of triangular lattice within a square matrix ({\bf i}) and again let the solid recover the square structure: ({\bf ii} and {\bf iii} as above)  under the influence of the same field as in {\bf a}. The particles are coloured according to the value of the local $\chi$ field. {\bf c.} The resulting $P(\chi)$ at the end of runs {\bf a} (red circles)  and {\bf b} (blue squares), the solid line is the harmonic theory prediction~\cite{sas3}. }
\end{center}
\vskip -.5cm
\end{figure*}  

We end by commenting on the mechanism by which the $S_T$ structure is stabilised. Consider the two crystal lattices the target, $S_T$, and the equilibrium structure that the solid prefers without $h_X$, viz. $S_E$. The lowest energy path from $S_T$ to $S_E$ (a) may involve non-affine displacements of particles within $\Omega$, or ``atomic shuffles" or (b) may be purely affine as is the case of structures connected by a group-subgroup relation~\cite{kaunat}. Possibility (a) is easy to analyse since, in this case, the term involving $h_X$ in (\ref{hamil}) increases the energy of $S_E$ relative to $S_T$ stabilising the latter; any small fluctuation taking $S_T \to S_E$ {\em always} cost energy. Possibility (b) is more subtle.

For example, both square and triangular crystals may be considered as special cases of a general {\it oblique} lattice and one can be obtained from the other by a purely affine transformation. Along this transformation path, which involves a bulk homogeneous strain, therefore, $h_X$ will {\em not} contribute. However, such an event is statistically unlikely except for extremely small systems. What is more likely is that a small patch of particles locally transforms to the $S_E$ structure creating $\chi$ at the interface. Within classical nucleation theory~\cite{CL} the free energy cost of such a patch of size $L_p$ is ${\cal F}_p = A \Delta {\cal F} L_p^2 + B h_X L_p$, where we ignore interfacial terms independent of $h_X$. Here $\Delta {\cal F}$ is the bulk free energy difference per unit area between the two lattices and $A$, $B$ are constants. This patch is stable only if it is larger than a critical size $L_p^* \sim -h_X/\Delta {\cal F}$ and costs interfacial energy $\sim h_X^2/\Delta {\cal F}$. $L_p^*$ can be large if $\Delta {\cal F}$ is small and can be made even larger by tuning $h_X$. If $L_p^* > L$, again, a finite $S_T$ crystal will be stable due to {\em the inability of the system to create a sufficiently low energy $S_E|S_T$ interface}.

It is possible to demonstrate both these mechanisms for the square to triangular transition. It is known that a mechanically unstable square lattice may decay, at nonzero temperatures into the stable triangular structure in many different ways~\cite{spring}. For example, (a) alternate rows of particles in the square lattice may shift by half a lattice spacing hence producing a distorted triangular lattice which subsequently equilibrates. On the other hand a patch of particles inside the square matrix may undergo a distortion to the triangular structure (b). Our stabilisation strategy should be able to recover the target square lattice $S_T$ from both ( a \& b) of these distortions.   

In Figs.~\ref{patch}{\bf a} and {\bf b} we demonstrate this explicitly by starting from initial (square) configurations of a $N = 2500$ GCM solid incorporating these two kinds of distortions (a) and (b) and equilibrating with $h_X = -2.0$, where the square lattice is stable. In the first case (Fig.~\ref{patch} {\bf a}), the local shuffles of particles cost energy and are quickly removed from the solid. In the second case, (Fig.~\ref{patch} {\bf b (i)}) we first create a large patch of particles with local triangular order using an inhomogeneous affine strain.  Next, we equilibrate the surrounding matrix keeping the particles within the patch immobile. This produces a mechanically relaxed (but high energy with large local $\chi$) interface between a square solid and a triangular inclusion. The constraint is then removed and the whole system equilibrated. Fig.~\ref{patch}{\bf b (iii)} shows that the sub-critical patch thus created disappears. As a check (Fig.~\ref{patch}{\bf c}) we ensure that $P(\chi)$ obtained from the equilibrated configurations again match theoretical predictions~\cite{sas1,sas3}. 


In conclusion, we present a proposal for producing crystalline templates for colloidal particles using dynamic, feedback controlled, laser traps. These templates are reconfigurable and can, in principle, stabilise any structure in any dimension, though they are probably much easier to implement in 2d. Unlike static templates, our method does not merely provide a restoring force to the target structure. All possible affine fluctuations about $S_T$ are allowed, preserving all symmetries of the crystal. Only non-affine excursions away from $S_T$ are {\em selectively} suppressed. We believe that using a similar strategy even inhomogeneous structures such as surfaces and interfaces of any specified orientation as well as random glassy configurations may be stabilised. We hope that our work motivates experimental work in this direction in the near future. 

 \begin{acknowledgments}
The authors thank J. Horbach,  P. Nath and S. Egelhaf for discussions. SS acknowledges support from the FP7-PEOPLE-2013-IRSES grant no: 612707, DIONICOS. 
\end{acknowledgments}
%
%


\end{document}